\def\comment#1{}

\newcommand{\beg}{\begin{eqnarray}}
\newcommand{\eee}{\end{eqnarray}}

\documentclass[prl,aps,amsfonts,amssymb,draft,floats,twocolumn]{revtex4}
\def\cm#1{}

\newcommand{\qq}{{\frac{\Psi_1^2}{m_1}}}
\newcommand{\ww}{{\frac{\Psi_2^2}{m_2}}}
\begin{document}
\title{ A marginal remark on massless neutral boson,  in   $U(1)\times U(1)$ superconductors}
\author{
Egor Babaev\thanks{Email: egor@teorfys.uu.se \  http://www.teorfys.uu.se/people/egor/   
}}
\address{
Institute for Theoretical Physics, Uppsala University 
Box 803, S-75108 Uppsala, Sweden 
}
\begin{abstract}
We make some remarks on massless neutral boson in two-flavour Abelian Higgs model. 
{\bf Note added:} {\it this remark was merged with journal version of cond-mat/0111192 
on a referee request.}
 \end{abstract}
\maketitle
\newcommand{\la}{\label}
\newcommand{\aaa}{\frac{2 e}{\hbar c}}
\newcommand{\Pfaff}{{\rm\, Pfaff}}
\newcommand{\kA}{{\tilde A}}
\newcommand{\G}{{\cal G}}
\newcommand{\cP}{{\cal P}}
\newcommand{\M}{{\cal M}}
\newcommand{\E}{{\cal E}}
\newcommand{\btd}{{\bigtriangledown}}
\newcommand{\W}{{\cal W}}
\newcommand{\X}{{\cal X}}
\renewcommand{\O}{{\cal O}}
\renewcommand{\d}{{\rm\, d}}
\newcommand{\bfi}{{\bf i}}
\newcommand{\e}{{\rm\, e}}
\newcommand{\bfx}{{\bf \vec x}}
\newcommand{\bfn}{{\bf \vec n}}
\newcommand{\bfE}{{\bf \vec E}}
\newcommand{\bfB}{{\bf \vec B}}
\newcommand{\bfv}{{\bf \vec v}}
\newcommand{\bfU}{{\bf \vec U}}
\newcommand{\ccc}{{\vec{\sf C}}}
\newcommand{\bfp}{{\bf \vec p}}
\newcommand{\f}{\frac}
\newcommand{\bfA}{{\bf \vec A}}
\newcommand{\non}{\nonumber}
\newcommand{\be}{\begin{equation}}
\newcommand{\ee}{\end{equation}}
\newcommand{\ba}{\begin{eqnarray}}
\newcommand{\ea}{\end{eqnarray}}
\newcommand{\bastar}{\begin{eqnarray*}}
\newcommand{\eastar}{\end{eqnarray*}}
\newcommand{\half}{{1 \over 2}}

\narrowtext
The two-gap superconductivity appears being a very interesting phenomenon
\cite{mult}- \cite{ashc1} both from experimental and
 theoretical points of view.

Also, the two-gap superconductivity 
was argued to occur in liquid metallic hydrogen and deuterium
\cite{ashc1}.

A  two-gap superconductor 
can be   described by a two-flavour (denoted by $\alpha =1,2$) 
Ginzburg-Landau  functional:
\beg
&&
F =\int d^3x \ \biggl[ \frac{1}{2m_1} \left| \left(\nabla +
i \e{\bf A}\right) \Psi_1 \right|^2 + 
\frac{1}{2m_2}  \left| \left( \nabla +
i e {\bf A}\right) \Psi_2 \right|^2 
\nonumber \\
&& 
+ { V} (|\Psi_{1,2}|^2) + \eta [\Psi_1^*\Psi_2+\Psi_2^*\Psi_1]
+ \frac{{\bf B}^2}{2}
\biggr]
\la{act}
\eee
where $\Psi_\alpha = |\Psi_\alpha|e^{i \phi_\alpha}$ and $
{V} (|\Psi_{1,2}|^2)=-b_\alpha|\Psi_\alpha|^2+ 
\frac{c_\alpha}{2}|\Psi_\alpha|^4
$.
The term $\eta [\Psi_1^*\Psi_2+\Psi_2^*\Psi_1]
=2\eta |\Psi_1\Psi_2|\cos(\phi_1-\phi_2) $ 
describes interband Josephson effect \cite{legg}.
GL functional may aslo contain other  terms 
which we drop in order to
keep this discussion
as simple as possible.

Much insight  into two gap superconductivity can be gained 
by the equivalence mapping between the two-gap
GL model and a version of Faddeev nonlinear 
$O(3)$-sigma model in interaction with a vector field
\cite{we} which is an extension of the model \cite{fadde}
which earlier was found being relevant 
 in particle physics  \cite{nature,cho}
(see also a remark \cite{tri}). Based on equivalence
mapping in  \cite{we} it was shown that 
two-band superconductors allow
knotted solitons characterized by a nontrivial Hopf invariant.
The existence of  self-stablized finite-length defects
is one of the unique features of two-gap superconductor
which, as it was found later has only a counterpart in 
ferromagnetic triplet superconductors
\cite{tri}.

In the London limit  \cite{frac}
in a simply-connected space  the model consists of
decoupled  massless neutral O(2) and charged O(2) Bose fields.
In \cite{frac} were discussed  several types of vortices
allowed in the London limit - in particular fractional
flux  vortices and composite one-flux-quantum vortices 
(for an extended discussion see a followup paper
\cite{new9}, see also remark \cite{volov}).
In this remark we make a connection
with  a different formalism 
to describe these vortices.

From (\ref{act}) follows the equation 
for supercurrent
\beg
{\bf J}& =& \f{i e}{2m_1}[\Psi_1^*\nabla \Psi_1 - \Psi_1\nabla\Psi_1^*]
+ \f{i e}{2m_2}[\Psi_2^*\nabla \Psi_2 - \Psi_2\nabla\Psi_2^*]
\nonumber \\&&
+e^2 {\bf A}\Bigg[ \f{\Psi^2_1}{m_1}+\f{\Psi_2^2}{m_2}\Bigg]
\label{j}
\eee

Insight into the physical roots of the presence in the system 
of a neutral massless boson can be gained by considering  simplest topological
defects in this system. Such a defect corresponds to the situation 
when only phase $\phi_1$ changes $2 \pi$ 
around the core while the phase of the second
condensate remains constant. In that case we 
have from (\ref{j}):
\beg
{\bf A} ={\bf J} \Biggl[ e^2 \Bigl(  \f{\Psi^2_1}{m_1}
+ \f{\Psi^2_2}{m_2} \Bigr) \Biggr]^{-1}+ 
\f{1}{e}\f{\qq}{\qq+ \ww} \nabla\phi_1
\eee
From this expression it follows \cite{frac} that 
this vortex carries the following fractional magnetic flux
\beg
\Phi=\oint {\bfA} d l=\f{\qq}{\qq+ \ww} \Phi_0,
\eee
where $\Phi_0=\f{2\pi}{e}$ is standard flux quantum.

Let us make some connections with
the results in \cite{frac} using more traditional
notations than that used in \cite{frac}.
Further exploring the solution for 
the vortex ($\Delta \phi_1=2\pi, \Delta \phi_2=0$)
we can write the vector potential 
as 
\be
{\bf A} = \f{{\bf r}\times {\bf e}_z}{|r|}|{\bf A}(r)|
\ee
where $r$ measures distance from the core 
and ${\bf e}_z$ is a unit vector pointing along the core.
The magnetic field is then given by
\be
|{\bf B}|=\f{1}{r} \f{d}{d r}(r |{\bf A}|)
\ee
The equation (\ref{j}) can then be rewritten
as 
\beg
-\f{d}{dr}\Bigl[\f{1}{r}\f{d}{d r}(r|{\bf A}|) \Bigr]
+\qq\Bigl[|{\bf A}|e^2-\f{e}{r}\Bigr] +\ww|{\bf A}|e^2 =0
\nonumber
\eee
For such a vortex the solution for vector potential is 
\beg
|{\bf A}|=\f{\qq}{\qq+\ww}\f{1}{er}-
\f{\qq}{\sqrt{\qq+\ww}}K_1\Bigg(e \sqrt{\qq+\ww} r\Bigg)
\label{A}
\eee
Indeed the 
magnetic field vanishes exponentially 
from the vortex core
at the characteristic length scale of magnetic
field penetration length $\lambda=\Big[e \sqrt{\qq+\ww}  \Big]^{-1}$
\cite{frac}: 
\be
|{\bf B}|=e\qq K_0\Bigg(e \sqrt{\qq+\ww} r\Bigg)
\ee
Also in contrast to the Abrikosov vortex \cite{aaa},
besides fractionalization  of magnetic flux 
such a vortex also features neutral vorticity. This, in particular, can be
seen by substituting the solution (\ref{A})
into (\ref{act}). At the length scales larger than
magnetic field penetration length from 
the vortex core it gives 
the following expression for the energy density:
\beg
F&=& \f{1}{2m_1}\Bigg|\Bigg( \nabla + i{\qq}\Bigg[{\sqrt{\qq+\ww}}\Bigg]^{-1}\f{1}{r}\Bigg)\Psi_1 \Bigg|^2
\nonumber \\
&+& 
\f{1}{2m_2}\Bigg|\Bigg(  i{\qq}\Bigg[{\sqrt{\qq+\ww}}\Bigg]^{-1}\f{1}{r}\Bigg)\Psi_2 \Bigg|^2
\label{nnn}
\eee
Thus the energy  per unit length of the
 vortex ($\Delta \phi_1=2\pi, \Delta \phi_2=0$)
 is divergent. This is due to the fact that 
such a topological configuration
necessarily induces in two-gap system a neutral superflow.
Indeed the expression (\ref{nnn})
is equivalent to the energy density 
in a {\it neutral} system with a vortex in
in a {\it neutral} phase field $\phi_1$
with effective stiffness $\qq \ww \Big[ \qq+\ww\Big]^{-1}$:
\be
F= \f{1}{2}
{\qq \ww}\Bigg[{\qq+\ww}\Bigg]^{-1}
| \nabla \e^{i \phi_1} |^2
\ee
Let us remark that this procedure is equivalent 
to that used in \cite{frac} with the only difference
that in \cite{frac} first the variables
were separated  into a neutral and a charged fields
(the London limit of the general procedure \cite{we}), 
and then there were found the solutions
for the vortices, 
while in the presentation in this paper  first
the solution found in \cite{frac} 
is substituted in (\ref{act})
and then the divergent part associated
with neutral vorticity extracted from both kinetic terms 
in (\ref{act}).

It displays in a transparent way  the roots 
of the presence in the system of a massless
neutral boson:   a topologically nontrivial
configuration  ($\Delta \phi_1=2\pi, \Delta \phi_2=0$),
besides current in the field ``1" also necessarily
induces  current in the component ``2".
Albeit in such a configuration there are no gradients of $\phi_2$
however the two condensates are not independent
but are connected by the vector potential.
Admixture of oppositely directed superflow of the component 
``2"  which necessarily accompanies such a vortex in two-gap
system, leads to situation when two superflows
partially compensate the induced by each other magnetic 
field which leads
to  the  existence of the effective neutral superflow
in the system.
 From (\ref{act})
it is seen that when we can not neglect the term
$2\eta |\Psi_1 \Psi_2|\cos(\phi_1-\phi_2)$ we deal with  sine-Gordon
vortices, the regime with finite $\eta$ is relevant for 
two-band superconductors like $MgB_2$
where two condensates are not independently conserved.

In conclusion: earlier it was shown that the model (\ref{act}) is dual 
to Faddeev model which consists of a unit
vector field with $O(3)$ symmetry
and a massive vector field which are coupled
by Faddeev term. In the London limit 
these variables decouple. But in a non-simply-connected space 
(zero of the order parameter in the vortex
core makes the space being non-simply-connected) 
even in the London limit these fields
remains topologically coupled. Which results in the fractionalization
of magnetic flux.
So in the London limit in a non-simply-connected space
the system may be viewed as topologically
coupled neutral and charged $O(2)$ bosons.
In this brief note we made some
explanatory remarks on the physical roots of
this effect.
  
The author is grateful to  Prof. G.E. Volovik, 
 Prof.  S. Girvin,  D.F. Agterberg 
and especially to D. Gorokhov, K. Zarembo and V. Cheianov
for  discussions and/or coments on this topic. 
It is also a great pleasure to thank  
Prof. Ludwig D. Faddeev, Prof.  Antti J. Niemi, 
for many general discussions of the  model (\ref{act}). 
This work has been supported by grant
 STINT IG2001-062 and the Swedish
Royal Academy of Science.

{\bf Notes added:} { 
We call reader's attention to  long range interacting pancake vortices
 and BKT transitions which have been considered in connection 
with High-$T_c$ layered superconductors in 
\cite{pankakes}, the author thanks G. Blatter for pointing that out.}
{ \bf This eprint was merged on a referee request with journal version of Ref. \cite{frac} }

\end{document}